\documentstyle[12pt]{article}
\newcommand{\beq}{\begin{equation}}
\newcommand{\eeq}{\end{equation}}
\newcommand{\bea}{\begin{eqnarray}}
\newcommand{\eea}{\end{eqnarray}}
\newlength{\bredde}
\def\slash#1{\settowidth{\bredde}{$#1$}\ifmmode\,\raisebox{.15ex}{/}
\hspace*{-\bredde} #1\else$\,\raisebox{.15ex}{/}\hspace*{-\bredde} #1$\fi}
\begin{document}
%\begin{Ntitlepage}
%\docnum{CERN--TH-7035/93}
%\vspace{1cm}

\title{\bf Collective Fields for QCD}

\author{{\sc P.H. Damgaard}\thanks{\em
Invited talk presented by P.H. Damgaard at the workshop on
``Quantum Field Theoretical Methods in High Energy Physics'',
Kyffh\"{a}user, Germany, Sept. 20-24, 1993. To appear in those
proceedings.}
            \\
CERN -- Geneva \\ ~~ \\ {\sc H.B. Nielsen} \\ The Niels Bohr Institute
\\ Blegdamsvej 17, DK-2100 Copenhagen, Denmark \\ ~~ \\ and \\ ~~ \\
{\sc R. Sollacher} \\ Gesellschaft f\"ur Schwerionenforschung GSI
mbH\\ P.O.Box 110552, D-64220 Darmstadt, Germany}

\maketitle
\vfill
\begin{abstract}
A gauge-symmetric approach to effective Lagrangians is described with
special emphasis on derivations of effective low-energy Lagrangians from
QCD. The examples we discuss are based on exact rewritings of cut-off
QCD in terms of new collective degrees of freedom. These cut-off
Lagrangians are thus ``effective'' in the sense that they explicitly
contain some of the physical long-distance degrees of freedom from the
outset.
\end{abstract}

%\vspace{15mm}

%\vspace{45mm}

\begin{flushleft}
CERN--TH-7035/93 \\ hep-ph/9310354 \\
\end{flushleft}
%\end{Ntitlepage}
%\thispagestyle{empty}
\vfill\eject
%\pagestyle{empty}
%\clearpage\mbox{}\clearpage

%\vfill
\newpage

It is well known that a description of physics in terms of given field
variables $\phi_i(x)$ and a corresponding action $S[\phi_i]$ is far from
being unique. There are numerous examples in quantum field theory where
exactly the same physical results can be obtained by entirely different
actions $\tilde{S}[\tilde{\phi}_i]$ in terms of different variables
$\tilde{\phi}_i$. This is a consequence of the {\em field redefinition
theorem}, which under very specific conditions ensures that $S$-matrix
elements remain unaltered under a change of basis of the fields (for a
partial list of references, see, e.g., ref. \cite{redef}). From a modern
path-integral point of view the field redefinition theorem may seem
almost trivial since it appears to express nothing but the possibility
of changing variables inside an integral, but there are in fact subtleties
at the two-loop level \cite{Gervais,AD1}. These stem from the fact that
a path integral is, after all, {\em not} just an ordinary integral. But
when these problems appear, and how they can be avoided, is by now well
understood.

More surprising is the fact that even such a highly non-trivial
example as (1+1)-dimensional {\em bosonization} \cite{boson} also can
be viewed as a genuine field redefinition \cite{AlfDam:90,us}. This is
most easily seen in the collective field formalism which is at the heart
of the effective Lagrangian method we will describe here. This technique
also reveals how bosonization (and fermionization) are only two extremes
of a continuum of equivalent field theory descriptions that contain, in
general, both bosons and fermions with non-trivial interactions. In this
sense, the massive Thirring model and the Sine-Gordon model are both
effective Lagrangian descriptions of the same physics. One is no more
fundamental than the other, but they may each have different domains in
which they are more simple -- more ``effective'' -- in describing
physics. In fact, and this turns out to be of crucial importance, both
can be viewed as particular {\em gauge fixings} of a ``higher''
gauge-symmetric theory that contains both bosons and fermions \cite{us}.

We are hereby implying that the concept of effective Lagrangians must
be broadened to include also the description of the same theory in
terms of more convenient variables. As an example, Lagrangians written
in terms of local fields which we can associate with physical states
should thus, in many instances, be far more ``effective'' than one
described by means of underlying fundamental fields.

Knowing that the gauge-symmetric collective field technique
\cite{HoKi,AlfDam:90} can be used to understand two-dimensional
bosonization (and actually much more besides) \cite{us}, it is an obvious
challenge to try to extend the same method to a full-fledged
four-dimensional
field theory such as QCD. The extent to which this is possible has recently
been discussed in ref. \cite{us1}, and the purpose of this talk is to
provide a short review of that work. We shall try to highlight both weak
and strong points of such an approach. In the course of our investigation
we came across work by Zaks and by Karchev and Slavnov
\cite{Slavnov} which implicitly contain related ideas. There are also clear
connections to the more conventional approach toward a derivation of
chiral Lagrangians from QCD \cite{chiral}. Related considerations can
be of relevance for the electroweak theory \cite{Bielefeld}; see also
ref. \cite{Bielefeld1}.

\vspace{0.7cm}

We begin by explaining the idea that {\em any} quantum field theory
described by an action $S[\phi_i]$ (and an associated measure in the
functional integral) can be considered as the gauge-fixed version of a
class of ``higher'' gauge-symmetric Lagrangians. To find such a higher
theory is not difficult. One first performs a field-enlarging
transformation
$G_a:~ \phi_i \to \phi'_i$, or $\phi_i = g_i(\phi',a)$, where $a_\alpha(x)$
are what we will call collective fields. They can at this stage be
considered
as parametrizing the transformation $G_a$. Under this transformation the
action should transform as a scalar, i.e. $S'[\phi',a] \equiv S[g_i(\phi',
a)]$. We will only consider (invertible) transformations
connected to the identity, which we take to occur at $a_\alpha(x) = 0$ (and
this can always be arranged). Under this transformation $G_a$, the measure
will pick up a Jacobian  $\det|M_{ij}(\phi',a)| \equiv \det|\delta
g_i(\phi',a)/\delta\phi'_j(x)|$. We next integrate over the transformation
fields $a_\alpha(x)$ in the functional integral. Since these fields can be
removed completely from both action and measure (by performing the inverse
transformation $G_a^{-1}$ on $\phi'_i$), integrating over $a_\alpha(x)$
does not affect any physics. It does, however, make the path integral
rather
ill-behaved due to the ``volume'' of the integration over $a_\alpha(x)$.
In the equivalent formulation in which $a_\alpha(x)$ is part of both
action and (new) measure, this manifests itself in the emergence of a new
{\em local gauge symmetry}:
\begin{eqnarray}
\delta \phi'_i(x) ~=~ - \int dy dz M^{-1}_{ij}(x,y)\frac{
\delta g_j(\phi'(y),a(y))}{\delta a_\alpha(z)}\Lambda_\alpha(z) ~,~~~
\delta a_\alpha(x) ~=~ \Lambda_\alpha(x) ~,
\end{eqnarray}
where $\Lambda_\alpha(x)$ is the gauge transformation parameter. When we
gauge-fix on the surface $a_\alpha(x) = 0$, we simply recover the theory
in the original formulation. Other gauges will in general correspond to
field redefinitions. This means that we can trade some of the original
fields $\phi_i(x)$ for some of the collective fields $a_\alpha(x)$.

The first question to answer is: Why do we choose to perform field
redefinitions using what appears to be such an elaborate route?
(First we extend the basis of fields by new collective fields, then we
immediately gauge-fix the same number of excessive degrees of freedom
away).
The reason is that this is an extremely efficient way of performing  field
redefinitions of arbitrary complexity. One {\em can} always obtain the same
redefinitions by direct changes of variables in the functional integral,
but it is not obvious how to find the corresponding direct
transformation. In contrast, using standard (BRST) gauge-fixing technology
this is accomplished in one simple step. Furthermore, {\em we need not
know} the precise form of the field-redefinition transformation; the change
of variables can be done in an entirely indirect manner through the
gauge fixing.
But it is important to stress that the
gauge symmetry (1) should {\em not} be given any physical meaning. It is
an artifact of the method.

\vspace{0.7cm}

Let us now try to apply this technique to QCD. By {\em collective}
excitations of QCD, we mean those colour-singlet bound states that can
appear in asymptotic states (or at least have a sensible finite lifetime).
These are the mesons, glueballs, baryons, ``hybrids'', etc. The simplest
collective fields of QCD are those describing the pseudoscalar bosons,
and it is those we will focus on here. We will thus set out to find the
closest we can get to ``bosonizing'' some of the fermionic QCD degrees of
freedom. This we will do first in the (flavour) abelian case
by trying to extract directly from the QCD
Lagrangian a field with the quantum numbers of the $\eta'$ meson.

Our starting point is a generating functional for QCD of the form
\begin{eqnarray}
{\cal Z}_{QCD} [V,A] &=& \int\! {\cal D} [\bar{\psi},\psi]{\cal
D}\mu[G]\; e^{-\int\! d^4x\; {\cal L}_{QCD}} \cr {\cal L}_{QCD} &=&
\bar{\psi}(x) (\slash{\partial} -i\slash{G}(x)
-i\slash{V}(x) -i \slash{A} (x)\gamma_5)\psi(x) +
\frac{1}{4g^2} tr G_{\mu\nu} (x)G_{\mu\nu} (x)~.
\label{eq:ZQCD}
\end{eqnarray}
Here $V_\mu(x)$ is an external vector source and $A_\mu(x)$ an
external axial vector source, both Abelian (diagonal in the $SU(N_f)$
flavour indices). The vector potential $G_{\mu}(x)$ is the usual gluon
field, here for convenience generalized to $SU(N_c)$, and
$G_{\mu\nu}(x)$ is the corresponding field strength tensor.
The usual $SU(N_c)$ colour gauge symmetry of course has to be
gauge-fixed in the standard manner, including also Yang-Mills
ghosts. For the moment, we simply include these Yang-Mills
gauge-fixing terms implicitly in the gluon measure ${\cal{D}}\mu[G]$.

There is nothing unphysical implied by the coupling to external vector
and axial vector sources; these sources only serve to define
appropriate Green functions through functional differentiation. They
are clearly not intrinsically part of QCD, and will eventually be set
equal to zero.  Nevertheless, they turn out to play a rather profound
r\^{o}le in the derivation of the effective Lagrangian. They would
also, of course, acquire a physical meaning if they were to include
the couplings of the electroweak interactions.
The $\gamma$-matrices are Hermitean and obey the usual Clifford algebra.

We next perform a field-enlarging transformation in which the fermion
fields are chirally rotated by a local (abelian) field $\theta(x)$.
It is essential that the field
transformation is made only in a {\em regularized} version of the QCD
generating functional.  A convenient consistent scheme in the fermion
sector is provided by a set of Pauli-Villars regulator
fields. These regulators of course {\em only}
regularize the fermionic sector of QCD. We
still need to regularize also the gluon sector of QCD in order to
interpret the resulting field-transformed Lagrangian as an effective
Lagrangian with an ultraviolet cut-off $\Lambda$.

After the field transformation we will get two pieces, one classical
from the variation of the action in eq.(\ref{eq:ZQCD}) under the local
chiral rotation, and one quantum mechanical from the change of the
fermionic measure \cite{Ball}. We rearrange it into an expansion in
decreasing powers of $\Lambda$, the ultraviolet cut-off:
\begin{eqnarray}
{\cal Z}_\Lambda [V,A] &=& \int\! {\cal D}_\Lambda [\bar{\chi},\chi
]{\cal D}\mu[G]\; e^{-\int\! d^4x\; {\cal L}'}
\cr {\cal L}' &=& \frac{1}{4g^2}tr G_{\mu\nu}G_{\mu\nu} +
\bar{\chi} (\slash{\partial} -i\slash{G} - i\slash{V}
-i\slash{A}\gamma_5 +i\slash{\partial}\theta \gamma_5) \chi
+ {\cal{L}}_{WZ} + {\cal{L}}_J
\label{eq:Znew}
\end{eqnarray}
where the last two terms arise from the Jacobian of the
transformation. The first part, the Wess-Zumino term, starts at
${\cal{O}}(\Lambda^0)$:
\beq
{\cal L}_{WZ} = -\frac{iN_f}{16\pi^2}
\theta\epsilon_{\mu\nu\rho\sigma}\biggl( tr G_{\mu\nu}
G_{\rho\sigma} - 4N_c \partial_\mu V_\nu \partial_\rho V_\sigma -
\frac{4N_c}{3} \partial_\mu A_\nu \partial_\rho A_\sigma \biggr) +
{\cal O}(\Lambda^{-2}),
\label{eq:LWZ}
\eeq
while the second part reads ${\cal{L}}_J = \Lambda^2{\cal{L}}_2 +
{\cal{L}}_0 + \frac{1}{\Lambda^2}{\cal{L}}_{-2} + \frac{1}{\Lambda^4}
{\cal{L}}_{-4} + \ldots$, with
\bea
{\cal L}_2 &=& \frac{N_f N_c \kappa_2}{4\pi^2} \biggl( A_\mu A_\mu
-(A_{\mu} - \partial_{\mu}\theta)(A_{\mu} -
\partial_{\mu}\theta)\biggr) \cr
{\cal L}_0 &=& \frac{N_f N_c}{24\pi^2} \biggl( \partial_\mu A_\nu
\partial_\mu A_\nu - \partial_\mu(A_\nu - \partial_{\nu}\theta)
\partial_\mu (A_\nu - \partial_{\nu}\theta) \cr &&
\qquad+ 2 \Bigl( A_\mu A_\mu \Bigr)^2 - 2\Bigl((A_\mu -
\partial_\mu \theta )(A_\mu -\partial_\mu\theta)\Bigr)^2 \biggr) \cr
{\cal L}_{-2} &=& \frac{N_f \kappa_{-2}}{48\pi^2}
\biggl(N_c \partial^2 A_\mu \partial^2 A_\mu - N_c \partial^2 (A_\mu -
\partial_\mu \theta) \partial^2 (A_\mu - \partial_\mu \theta ) \cr
&& \qquad + \Bigl( A_\mu A_\mu - (A_\mu -\partial_\mu \theta )( A_\mu
-\partial_\mu \theta ) \Bigr)\;tr_c G_{\nu\rho} G_{\nu\rho} + {\cal
O}(A_\mu^4)~.
\biggr)
\label{eq:J+5}
\eea
We list only the first three terms; the whole expansion in increasing
powers of inverse cut-off can be computed following the technique
described in ref. \cite{Ball}.
The coefficients $\kappa_2$ in eq.  (\ref{eq:J+5})
are regularization-scheme dependent constants \cite{us1}.

When we next integrate over the collective fields in the path
integral, a chiral gauge symmetry appears \cite{AlfDam:90}:
\bea
\chi (x) &\to& e^{i\alpha (x) \gamma_5} \chi (x) \cr
\bar{\chi} (x) &\to& \bar{\chi} (x) e^{i\alpha (x) \gamma_5} \cr
\theta (x) &\to& \theta (x) - \alpha (x).
\label{eq:sym5}
\eea

The transformed action (\ref{eq:Znew})
already contains a kinetic energy piece for $\theta(x)$:
\bea
{\cal{L}'} &=&
\frac{1}{4g^2}tr G_{\mu\nu}G_{\mu\nu}
+ \bar{\chi} (\slash{\partial} -i\slash{G} - i\slash{V}
-i\slash{A}\gamma_5 +i\slash{\partial}\theta \gamma_5) \chi
\cr & & +\frac{N_f}{2} \partial_\mu \theta f^2 \partial_\mu \theta -
N_f A_\mu f^2 \partial_\mu \theta -\frac{N_f N_c}{12\pi^2}
\biggl(\Bigl((A_\mu -
\partial_\mu \theta )(A_\mu
-\partial_\mu \theta ) \Bigr)^2 -
\Bigl( A_\mu A_\mu \Bigr)^2 \biggr) \cr
&& - \theta \frac{iN_f}{16\pi^2} \;\epsilon_{\mu\nu\rho\sigma}
\biggl( tr_c G_{\mu\nu} G_{\rho\sigma} - 4N_c \partial_\mu V_\nu
\partial_\rho V_\sigma -\frac{4N_c}{3} \partial_\mu A_\nu
\partial_\rho A_\sigma \biggr)
 + {\cal O} (\Lambda^{-2})~,
\label{eq:L'}
\eea
where $f^2$ is an operator:
\beq
f^2 = -\frac{N_c \kappa_2 \Lambda^2}{2\pi^2} + \frac{N_c}{12\pi^2}
\partial^2 - \frac{N_c \kappa_{-2}}{24\pi^2 \Lambda^2} \partial^2
\partial^2 - \frac{\kappa_{-2}}{24 \pi^2
\Lambda^2} \; tr_c  G_{\nu\rho} G_{\nu\rho} + \ldots
\label{eq:f2}
\eeq
The dots denote higher-order gluonic terms, derivatives and
combinations of them divided by suitable powers of $\Lambda$.
The implied expansion in inverse powers of the
ultraviolet cut-off $\Lambda$ is somewhat deceptive, since these powers
are often  multiplying operators of increasing dimension.
This can lead to compensating factors of $\Lambda$ in the numerators,
thus putting the validity of the expansion in jeopardy. The only genuine
expansion parameter will then be (low) momentum, or
inverse powers of a large number of colours $N_c$.

Suppose we introduce
a pseudoscalar field $\eta_0$ with the canonical dimension of $mass$
as $\theta = \eta_0/\sqrt{N_f}f_0$,
and view  $f_0$ as a bare coupling. Identifying $f_0^2$ with the
leading term in (\ref{eq:f2}),$
f_0^2 = -N_c \kappa_2 \Lambda^2/2\pi^2$,
leads to a Lagrangian for a pseudoscalar field:
\beq
{\cal L}_{\eta_0} = \frac{1}{2} \partial_\mu \eta_0 \partial_\mu
\eta_0 - A_\mu \sqrt{N_f} f_0 \partial_\mu \eta_0 + \ldots
\eeq
The dots denote higher derivative terms, gluonic terms and self
interactions of the pseudoscalar field. Note that $f_0$ is both
cut-off dependent and scheme dependent.
Let us now look at the higher derivative terms. In a perturbative
sense, the propagator for the field $\eta_0$ can be derived from the
bilinear part of ${\cal L}'$.
It is a higher-derivative (or essentially Pauli-Villars)
{\em regularized} bosonic propagator with a
regulator mass proportional to $\Lambda^2$.
There is probably a simple reason for this: we are
throughout performing field transformations within a {\em regularized}
fermionic path integral. Even after a series of field-enlarging
transformations
(and the required gauge fixing of the new local symmetry) of such a
form that we end up with new propagating fields, the generating
functional is still ultraviolet regularized.

As explained above,
useful forms of the effective Lagrangian are derived by judicious
choices of gauge fixing. Whereas the gauge $\theta(x) = 0$ trivially
gives us back cut-off QCD in its original formulation, almost all
other gauge choices that remove part of the QCD degrees of freedom
will lead to non-trivial effective Lagrangians. We base the
gauge fixing it on
the axial singlet current as given
by a functional derivative with respect to $A_\mu$ at $A_\mu =0$:
\beq
i\langle J_{\mu}^5\rangle =
i\langle \bar{\psi} \gamma_\mu \gamma_5 \psi \rangle = i\langle
\bar{\chi} \gamma_\mu \gamma_5 \chi \rangle + N_f f^2
\partial_\mu \theta + \ldots
\label{eq:trAxCur}
\eeq
The additional terms represented by dots are at least of third order
in $\theta(x)$. The whole expression is of course gauge invariant,
but the individual components on the r.h.s. are not.
We now choose a gauge-fixing
function $\Phi$ by
\beq
\Phi =  i\frac{\partial_\mu}{N_f f_0^2\partial^2}
\bar{\chi} \gamma_\mu \gamma_5\chi ~.
\label{eq:Phi'}
\eeq
When we implement $\Phi$ as a delta-function constraint in the path
integral, we must be careful that it has correct
transformation properties.  Under a global chiral rotation,
$
\bar{\chi}(x) \to \bar{\chi}(x)e^{i\alpha\gamma_5} ~,~
\chi(x) \to e^{i\alpha\gamma_5}\chi(x),
{}~\bar{\chi}\gamma_{\mu}\gamma_5\chi$ remains {\em
classically} invariant but quantum mechanically it shifts due to the
chiral anomaly:
eq. (\ref{eq:L'}):
\beq
i\frac{\partial_\mu}{N_f f_0^2\partial^2}
\bar{\chi} \gamma_\mu \gamma_5\chi \to
i\frac{\partial_\mu}{N_f f_0^2\partial^2}
\bar{\chi} \gamma_\mu \gamma_5\chi + \alpha ~.
\eeq
The action does not remain invariant either,
but shifts due to the axial anomaly:
\beq
S' = \int\! d^4x\; {\cal{L}}' \to S' - 2i N_f \alpha \int\! d^4x
\frac{1}{32\pi^2}\; \epsilon_{\mu\nu\rho\sigma} tr_c G_{\mu\nu}
G_{\rho\sigma} ~.
\eeq
Assuming that we sum only over integer winding numbers, the action
does, however, remain invariant under constant chiral rotations of the
form $\alpha = n\pi/N_f$. This means that also $\theta(x)$ is only
globally defined modulo $\pi/N_f$.
The gauge-fixing constraint must respect the above
periodicity property; there must, even in the gauge-fixed path
integral, be no distinction between $\theta(x)$ and $\theta(x) +
n\pi/N_f$. If we choose a $\delta$-function constraint to implement
the gauge choice, this $\delta$-function must then necessarily be {\em
globally periodic}.
Representing it by an auxiliary field $b(x)$, the
gauge-fixing function then provides a few new terms in the
action. But they will in general modify
the relevant chiral Jacobian, and for consistency new terms must
be added to the action to compensate for this. This leads to a rather
involved procedure of gauge fixing,
and we give here a simpler derivation.
We can consider
the above gauge-fixing function as derived from a constraint in the
{\em original} representation of QCD of the form
\beq
\Phi' =  i\frac{\partial_\mu}{N_f f_0^2\partial^2}
\bar{\psi} \gamma_\mu \gamma_5\psi + \theta ~.
\eeq
We then introduce the $\delta$-function constraint as
\bea
\delta(\Phi')
&=& \int\! {\cal{D}}[b]\; \exp\left[-\int\!  d^4x\;
\left(\frac{i} {N_f f_0^2}B_{\mu}(x)J^5_{\mu}(x) +
\theta(x)\partial_{\mu}B_{\mu}(x)
\right)\right] ~,
\eea
where the axial vector field $B_{\mu}(x)$ is defined by
\beq
B_{\mu}(x) \equiv \int\! d^4y d^4z\; b(y)\partial^{-2}_{(y-z)}
\partial_{\mu}^{(z)}\delta(z-x) ~,
\eeq
i.e.,$b(x) = -\partial_{\mu}B_{\mu}(x)$. Global periodicity of the
$\delta$-function means that $b$ is constrained,
\beq
\int\! d^4x\; b(x) = -\int\! d^4x\; \partial_{\mu}B_{\mu} = ik N_f ~,
\eeq
where $k$ is an arbitrary integer. This global constraint means that
$b(x)$ (or $\partial_{\mu}B_{\mu}$) share certain properties with
topologically non-trivial fields.

After gauge-fixing the only remnant of the $U(1)$ axial gauge symmetry
is the BRST symmetry
\begin{eqnarray}
\delta \bar{\chi}(x) &~=~& i\bar{\chi}(x)\gamma_5c(x) \cr
\delta \chi(x) &~=~& -ic(x)\gamma_5\chi(x) \cr
\delta \theta(x) &~=~& -c(x) \cr
\delta \bar{c}(x) &~=~& b(x)~,
\end{eqnarray}
and $\delta c(x)= \delta b(x) = 0$.
The ghost term is trivial in the present case, being just $\bar{c}c$..

It is convenient to choose a slightly different gauge which will only
affect higher order correlation functions of our gauge fixing
expression $\Phi'$. We simply choose to add BRST-invariant terms
involving powers of $b(x)$ such
that the gauge fixed Lagrangian looks like
\bea
{\cal L}'' &=& \bar{\chi} \biggl( \slash{\partial} -i\slash{G}
-i\slash{V} -i\Bigl(\slash{A} -
\frac{1}{N_f f_0^2}\slash{B} -\slash{\partial} \theta
\Bigr) \gamma_5\biggr) \chi + \bar{c} c + {\cal L}_{YM}  \cr
&& +\frac{N_f f_0^2}{2} \partial_\mu \theta
\partial_\mu \theta - N_f f_0^2 A_\mu \partial_\mu \theta \cr
&& -\frac{N_f N_c}{12\pi^2} \biggl(\Bigl((A_\mu -\frac{1}{N_f f_0^2}
B_\mu - \partial_\mu \theta )(A_\mu -\frac{1}{N_f f_0^2} B_\mu
-\partial_\mu \theta ) \Bigr)^2 - \Bigl( A_\mu A_\mu
\Bigr)^2 \biggr) \cr
&& - \theta \frac{iN_f}{16\pi^2} \;\epsilon_{\mu\nu\rho\sigma} \biggl(
tr_c G_{\mu\nu} G_{\rho\sigma} - 4N_c \partial_\mu V_\nu \partial_\rho
V_\sigma -\frac{4N_c}{3} \partial_\mu A_\nu \partial_\rho A_\sigma
\biggr) + {\cal O} (\Lambda^{-2})~.
\label{eq:LDel1}
\eea

In order to view (\ref{eq:LDel1}) as an {\em effective Lagrangian}, we need
additional input. The obvious choice would be to identify the
$\theta$-field with the flavour-singlet pseudoscalar field of the
$\eta'$ meson, in appropriate units. Certainly, eq.  (\ref{eq:LDel1})
gives the correct QCD action for describing the low-momentum dynamics of
the composite operator $J^5_{\mu}(x) =
i\bar{\psi}\gamma_{\mu}\gamma_5\psi(x)$ of the original quark fields.
Taking one partial derivative, we can
equally well describe $\partial_{\mu}J^5_{\mu}(x)$, which is a
non-zero operator due to the chiral anomaly. It has quantum numbers
$J^{PC} = 0^{-+}$, and is a singlet under flavour.  As such, it
should have a non-vanishing overlap with the physical $\eta'$
meson. For example, if we were able to compute the long-distance
fall-off of the corresponding two-point correlation function, this
should provide us with the mass of the lowest-lying state of these
quantum numbers. By definition, this is the mass of the $\eta'$ meson.

Going back to
eq. (\ref{eq:ZQCD}), we note that the connected 2-point function
of $\partial_{\mu}J^5_{\mu}(x)$ can be obtained by differentiating
twice with respect to a pseudoscalar source $\sigma(x)$ defined by
splitting $A_{\mu} = \partial_{\mu}\sigma(x) + A_\mu^T$ into a
longitudinal and a transverse part. Shifting $B_\mu$
\beq
B_\mu (x) \to B_\mu (x) + N_f f_0^2 \partial_\mu \theta(x) - N_f f_0^2
\partial_\mu
\sigma (x)
\label{eq:shift}
\eeq
leads to a Lagrangian
\begin{eqnarray}
{\cal L}''' &=& \bar{\chi}\biggl(\slash{\partial} -i\slash{G}
-i\Bigl(\slash{A}^T - \frac{1}{N_f f_0^2}\slash{B}
\Bigr)\gamma_5 \biggr)
\chi + {\cal L}_{ghost} + {\cal L}_{YM}\cr
&& -\frac{N_f N_c}{12\pi^2} \biggl(\bigl((A_\mu^T - \frac{1}{N_f
f_0^2} B_\mu ) (A_\mu^T -\frac{1}{N_f f_0^2} B_\mu ) \bigr)^2 - \bigl(
A_\mu A_\mu \bigr)^2
\biggr) \cr
&& +\frac{N_f f_0^2}{2}
\partial_\mu \theta \partial_\mu \theta - N_f f_0^2 \partial_\mu \sigma
\partial_\mu \theta \cr
&& - \theta \frac{iN_f}{16\pi^2} \;\epsilon_{\mu\nu\rho\sigma}
\biggl( tr_c G_{\mu\nu} G_{\rho\sigma} - 4 N_c\partial_\mu
V_\nu \partial_\rho V_\sigma -\frac{4N_c}{3} \partial_\mu A^T_\nu
\partial_\rho A^T_\sigma \biggr)
+ {\cal O} (\Lambda^{-2})
\label{eq:L'''}
\end{eqnarray}
Apart from contact terms only a linear coupling of $\sigma$ to
$\theta$ is left. The remaining part of ${\cal O} (\Lambda^{-2})$ is
also independent of $\theta$ because it contains $\theta$ only in the
combination $B_\mu + N_f f_0^2 \partial_\mu \theta$; after the shift
(\ref{eq:shift}) $\theta$ disappears from these terms.

We can now derive some exact Ward identities, setting the external
sources to zero: The original anomalous Ward identity
\beq
\partial_\mu \langle J_{5\mu} \rangle = i\partial_\mu \langle \bar{\psi}
\gamma_\mu \gamma_5 \psi \rangle = -i \frac{N_f}{16\pi^2} \langle
G\tilde{G} \rangle + {\cal O}(\Lambda^{-2})
\eeq
is now just the equation of motion for the field $\theta$:
\beq
f_0^2 \partial^2 \langle \theta \rangle = -i
\frac{1}{16\pi^2}
\langle G\tilde{G} \rangle +{\cal O} (\Lambda^{-2}) ~.
\eeq
Analogously, we find for the 2-point function
in the original QCD representation:
\bea
\langle \partial_\mu J_{5\mu} (x) \partial_\nu J_{5\nu} (y) \rangle = -
\left(\frac{N_f}{16\pi^2} \right)^2 \langle G\tilde{G} (x) G\tilde{G} (y)
\rangle - N_f f_0^2 \partial^2\delta (x-y) +{\cal O} (\Lambda^{-2})~~.
\label{eq:chWI2}
\eea
The same identity can
be derived from a simple infinitesimal shift of $\theta$ to
second order:
\bea
N_f^2f_0^4 \langle \partial^2 \theta (x) \partial^2\theta (y))
\rangle = - \left( \frac{N_f}{16\pi^2} \right)^2 \langle
G\tilde{G} (x) G\tilde{G} (y) \rangle - N_f f_0^2 \partial^2
\delta (x-y) +{\cal O}(\Lambda^{-2})
\label{eq:2pt}
\eea
This illustrates that for relevant Green functions our
gauge identifies as {\em operators}
\beq
\partial_\mu J_{5\mu} \sim N_f f_0^2 \partial^2 \theta ~.
\eeq
The gauge-fixing procedure presented above thus amounts to introducing
explicitly, at the Lagrangian level, an ``interpolating" field
according to the relation (26).

Just in order to illustrate how non-trivial results can be extracted from
the effective Lagrangian (21), let us consider a very crude approximation,
the cumulant expansion.
In this framework we can integrate out all fields in (\ref{eq:L'''})
except $\theta$ to arrive at an effective Lagrangian
\beq
{\cal L}_{eff} = \frac{F^2_0}{2} \partial_\mu \theta
\partial_\mu \theta +
\frac{F_0^2 M_0^2}{2} \theta^2 + \ldots
\eeq
The dots denote higher derivative terms and self-interactions of order
$\theta^3$. The parameters $F_0$ and $M_0$ are defined through
\beq
F_0^2 M_0^2 = \int\! d^4x\; \left\langle \frac{N_f}{16\pi^2}
G\tilde{G} (x)
\frac{N_f}{16\pi^2} G\tilde{G} (0) \right\rangle_{trunc}
\label{eq:M0}
\eeq
and
\beq
F_0^2 = N_f f_0^2 - \int\! d^4x\; \frac{x^2}{8} \left\langle
\frac{N_f}{16\pi^2} G\tilde{G} (x) \frac{N_f}{16\pi^2} G\tilde{G} (0)
\right\rangle_{trunc}~~.
\label{eq:F0}
\eeq

The expression for $F_0^2 M_0^2$ in ({\ref{eq:M0}) looks similar to
the one derived by Witten
\cite{Witten} and Veneziano \cite{Veneziano} in the limit
$N_c \to \infty$. But
the expectation values $\langle \ldots \rangle_{trunc}$ have to be
taken with respect to a ``truncated'' version of QCD:
\bea
{\cal L}_{trunc} = \bar{\chi}\biggl(\slash{\partial} -i\slash{G}
+i\frac{1}{N_f f_0^2} \slash{B}\gamma_5 \biggr)\chi + {\cal L}_{ghost}
+ {\cal L}_{YM} -\frac{ N_c}{12\pi^2 N_f^3 f_0^8} \biggl( B_\mu
B_\mu \biggr)^2 + {\cal O} (\Lambda^{-2}).
\eea

At first glance one would argue that the topological susceptibility
has to be zero in such a theory because of the massless quarks. If
this were true, then our field $\theta$ would be massless. Indeed, for
the full theory we can derive from the Ward identities (\ref{eq:2pt})
that the r.h.s. of eq. (28) vanishes.
This is a well-known result in massless QCD. But if we make an analogous
step in the {\em truncated} theory, we get instead
\bea
\int\! d^4x\; \langle \partial_\mu B_\mu (x) \; \partial_\nu B_\nu (0)
\rangle_{trunc} &=& - \biggl(\frac{N_f}{16\pi^2}\biggr)^2 \int\!
d^4x\; \langle G\tilde{G} (x) G\tilde{G} (0)\rangle_{trunc} +
{\cal O} (\Lambda^{-2})~~.
\eea
It is the constraint  (17) that
suggests a non-vanishing topological susceptibility in the truncated
theory, and thus a non-zero mass for the field $\theta$. Without
this constraint, one could decouple the $B_{\mu}$-field from the
fermions through a chiral rotation \cite{Bardakci}, and the standard proof
of a vanishing
topological susceptibility in the theory with massless quarks would
go through.

Witten has argued \cite{Witten} that the vanishing of the
topological susceptibility in massless QCD can be considered as a
cancellation of the pure gluonic part by a contribution from the
$\eta'$ meson.  From our point of view, the possibility of a non-zero
topological
susceptibility in the {\em truncated} QCD arises due to the gauge
constraint, which precisely can be thought of as
removing the $\eta'$-part of the topological susceptibility.

One obvious difficulty with the present approach is that
everything is expressed in terms of bare parameters in the cut-off
theory. Being explicitly cut-off dependent, it is also scheme
dependent. The whole set of effective one-loop interactions between
the bosonic collective fields and left-over QCD degrees of freedom
indeed follow directly from the Pauli-Villars regulator fields.  This
is just as in the solvable case of two dimensions \cite{us}.
As it stands, the unrenormalized theory has, with
massless quarks, only one mass scale: that of the cut-off $\Lambda$.
This means that all dimensionful couplings in the effective theory are
given by powers of this ultraviolet cut-off. In the renormalized
theory this cut-off becomes replaced by a physical mass scale,
essentially
$\Lambda_{QCD}$. In the end, if one integrates out all
gluonic and quark degrees of freedom and leaves only the collective
fields, the physical couplings are directly related to gluonic and
fermionic correlators, moments thereof, and condensates. The precise
relationship between the couplings of the collective field Lagrangian
and these vacuum expectation values will be of roughly the kind
discussed in the case of the Witten-Veneziano relation above.
The fact that physical
couplings will be related to these Green functions is also evident if
we return to the definition of the gauge-fixing condition
(\ref{eq:Phi'}).  The term containing $f_0^2$
should in fact depend on $f^2$ with contributions also from gluonic
fields, and a full treatment should incorporate the effect of
integrating out the gluonic degrees of freedom. Intuitively, one
expects that one major effect of such a renormalization program is to
relate $f^2$ to gluonic condensates.

\vspace{0.7cm}

Finally, we shall briefly outline the generalization of the
present effective Lagrangian technique to the case of the $SU(N_f)$
pseudoscalar multiplet. In contrast with the $U(1)$ case discussed
above, the flavour non-singlet axial currents are exactly conserved in
the limit of massless quarks.
How do we introduce the appropriate collective fields for this
non-Abelian (flavoured) case? As before, the main input is the choice
of quantum numbers we wish to describe. We then start again with a
generating functional of QCD, eq.(2),
where we now take external sources $V_\mu$ and $A_\mu$ to be
elements of $SU(N_f)$. It is now most
convenient to introduce collective fields $\theta (x)$ by, $e.g.$,
purely left-handed transformations:
\beq
q_L(x) = e^{2i\theta (x)} \chi_L (x)~ ,~\bar{q}_L (x) =
\bar{\chi}_L (x) e^{-2i\theta (x)}
\label{eq:trL}
\eeq
$i.e.$ local phase transformations acting only on the left-handed
spinors,$
q_L = P_+ q ~ ,~\bar{q}_L = \bar{q} P_- ~ ,~ P_\pm =
\frac{1}{2} (1\pm \gamma_5 ).$
Also, define $
L_\mu = V_\mu + A_\mu ~,~ R_\mu = V_\mu - A_\mu$.

The transformation (\ref{eq:trL}) causes a change of the regularized
fermionic functional integral measure due to its handedness. In order
to calculate the corresponding contribution to the Lagrangian, we again
use Pauli-Villars regularization:
\begin{eqnarray}
{\cal Z}_\Lambda [V,A] &=& \int\! {\cal D}_\Lambda [\bar{\chi},\chi ]
d\mu [G]\; e^{-\int\! d^4x\; {\cal L}'} \cr {\cal L}' &=& \bar{\chi}
\gamma_\mu (\partial_\mu -iG_\mu -iL_\mu^\theta P_+ -i R_\mu P_- )
\chi + {\cal L}_J + {\cal L}_{WZ} + {\cal L}_{YM}~,
\label{eq:ZnewL}
\end{eqnarray}
where, with $U(x) = e^{2i\theta (x)}, ~L_\mu$ is modified to
$L_\mu^\theta = U^\dagger L_\mu U + iU^\dagger \partial_\mu U~.$

The positive parity part can, as in the abelian case, be ordered as an
expansion in inverse powers of the ultraviolet cut-off $\Lambda$.
The first three terms are given by
\bea
{\cal L}_2 &=& \frac{N_c\kappa_2}{4\pi^2} tr_f A_\mu^{(s)} A_\mu^{(s)}
\mid_{s=1}^0 \cr {\cal L}_0 &=&
\frac{N_c}{8\pi^2} tr_f \biggl(-i F_{\mu\nu}^{(s)}
[A_\mu^{(s)} ,A_\nu^{(s)} ] + \frac{1}{3} D_\mu^{(s)} A_\nu^{(s)}
D_\mu^{(s)} A_\nu^{(s)} - \frac{2}{3} (A_\mu^{(s)} A_\mu^{(s)} )^2 \cr
&&\qquad + \frac{4}{3} A_\mu^{(s)} A_\nu^{(s)} A_\mu^{(s)} A_\nu^{(s)}
\biggr)
\mid_{s=1}^0\cr
{\cal L}_{-2} &=& \frac{\kappa_{-2}}{48\pi^2} tr_f
\biggl( N_c \partial^2 A_\mu^{(s)} \partial^2 A_\mu^{(s)} +
 A_\mu^{(s)} A_\mu^{(s)} tr_c G_{\nu\rho} G_{\nu\rho} + \ldots
\biggr)\mid_{s=1}^0~.
\label{eq:J+L}
\eea
The terms omitted in ${\cal L}_{-2}$ and denoted by dots are
at least fourth order in $A_\mu^{(s)}$ and $V_\mu^{(s)}$.

The leading term of the negative parity part is the integrated
Bardeen-anomaly:
\bea
{\cal L}_{WZ} &=& \frac{i}{16\pi^2} \int_{1}^{0} ds
\;\epsilon_{\mu\nu\rho\sigma}\; tr_f tr_c \theta \biggl(
F_{\mu\nu}^{(s)} F_{\rho\sigma}^{(s)} + \frac{1}{3} A_{\mu\nu}^{(s)}
A_{\rho\sigma}^{(s)}+ \frac{32}{3} A_\mu^{(s)} A_\nu^{(s)} A_\rho^{(s)}
A_\sigma^{(s)} \cr &&\qquad +
\frac{8i}{3} ( F_{\mu\nu}^{(s)} A_\rho^{(s)} A_\sigma^{(s)}
+ A_\mu^{(s)} F_{\nu\rho}^{(s)} A_\sigma^{(s)} + A_\mu^{(s)}
A_\nu^{(s)} F_{\rho\sigma}^{(s)} )
 \biggr) + {\cal O} (\Lambda^{-2})~.
\label{eq:J-L}
\eea
The covariant derivatives and field strength tensors appearing in
eqs.  (\ref{eq:J+L}) and (\ref{eq:J-L}) are defined as $
{\cal D}_\mu A_\nu = \partial_\mu A_\nu -i [V_\mu ,A_\nu ]
,~ A_{\mu\nu} = {\cal D}_{[\mu} A_{\nu]} ,~ F_{\mu\nu}
= \partial_{[\mu} V_{\nu]} - i[V_\mu , V_\nu ] - i[A_\mu ,A_\nu ]$,
and the transformed fields appearing in (\ref{eq:J+L}) and
(\ref{eq:J-L}) as
\bea
V_\mu^{(s)} &=& \frac{1}{2} \biggl( e^{-2is\theta} L_\mu e^{2is\theta}
+ i e^{-2is\theta} \partial_\mu e^{2is\theta} + R_\mu \biggr) \cr
A_\mu^{(s)} &=& \frac{1}{2} \biggl( e^{-2is\theta} L_\mu e^{2is\theta}
+ i e^{-2is\theta}
\partial_\mu e^{2is\theta} - R_\mu \biggr)~~.
\label{eq:AVs}
\eea
The parameter $s$ ranging from $0$ to $1$ thus defines a continuous
transformation which, for $s=1$, coincides with (\ref{eq:trL}).

When we now integrate
over the invariant Haar measure $\int {\cal D}[U]$, a new local
non-Abelian gauge symmetry emerges:
\begin{eqnarray}
\chi_L (x) &\to& e^{2i\alpha (x)} \chi_L (x)\cr
\bar{\chi}_L (x) &\to& \bar{\chi}_L (x) e^{-2i\alpha (x)} \cr
U(x) &\to& U(x) e^{-2i\alpha (x)} ~.
\label{eq:symL}
\end{eqnarray}

As in the flavour-singlet case, the crucial next step is the choice of
gauge fixing. While we have a lot of freedom avaliable in choosing the
gauge-fixing function, it is not immediately obvious what will lead to
useful representations. In that respect the abelian case is far better
under control, since we there can rely on experience gained in the
solvable two-dimensional case. To achieve the same amount of
simplification in this non-Abelian case seems to require that we know
how to perform the ``smooth''
analogue \cite{us} of non-Abelian bosonization. Still, even without
explicitly specifying the non-Abelian gauge-fixing function, it is clear
that the final result will be closely related to what has become known as
the constituent chiral quark model of Manohar and Georgi \cite{MaGe}.
This is a chiral Lagrangian coupled to remnant quark degrees of freedom
(in our language the $\chi$-fields) and the gluons. These couplings are,
apart from those induced by the gauge fixing, entirely specified by QCD
in our approach. But to really analyze the resulting effective Lagrangian
requires that we decide on a useful gauge-fixing function. Work in that
direction is still in progress.

\noindent
{\sc Acknowledgement:} ~ The work of P.H.D. and H.B.N. has been partially
supported by Norfa grant no. 93.15.078/00.

\bibliographystyle{unsrt}

\end{document}